# New observations of the gas cloud G2 in the Galactic Center

S Gillessen[1], R Genzel[1,2], T K Fritz[1], F Eisenhauer[1], O Pfuhl[1], T Ott[1], J Cuadra[3], M Schartmann[4,1], and A Burkert[4,1]

[1]Max-Planck-Institute for Extraterrestrial Physics, Giessenbachstraße, 85748 Garching, Germany
[2]Departments of Physics and Astronomy, Le Conte Hall, University of California, 94720 Berkeley, USA
[3]Departamento de Astronomía y Astrofísica, P. Pontificia Universidad Católica de Chile, Vicuña Mackenna 4860, 7820436 Macul, Santiago, Chile
[4]Universitätssternwarte der Ludwig-Maximilians-Universität, Scheinerstr. 1, D-81679 München, Germany

We present new observations of the recently discovered gas cloud G2 currently falling towards the massive black hole in the Galactic Center. The new data confirm that G2 is on a highly elliptical orbit with a predicted pericenter passage mid 2013. The updated orbit has an even larger eccentricity of 0.966, an epoch of pericenter two months later than estimated before, and a nominal minimum distance of 2200 Schwarzschild radii only. The velocity gradient of G2 has developed further to 600 km/s FWHM in summer 2012. We also detect the tail of similar total flux and on the same orbit as G2 along the trajectory at high significance. No hydrodynamic effects are detected yet, since the simple model of a tidally shearing gas cloud still describes the data very well. The flux of G2 has not changed by more than 10% between 2008 and 2012, disfavoring models where additional gas from a reservoir is released to the disrupting diffuse gas component.

## 1. Introduction

Recently, Gillessen et al. (2012) reported the discovery of a gas cloud (nicknamed G2 by Burkert et al. 2012) with a mass of ≈ 3 Earth masses in the Galactic Center that is on a nearly radial orbit falling towards Sgr A*. The predicted time of closest approach is in mid 2013. G2 can be observed in L-band (3.8$\mu$m) images and spectroscopically in atomic recombination lines of hydrogen and helium. The L-band emission originates from a relatively cool dust component (≈600K), while the emission lines are plausibly excited by the UV light of the surrounding star cluster. Most spectacular is the integral field spectroscopy between 2004 and 2011 showing that G2 developed a velocity gradient along its orbit - G2 is being tidally stretched. The multi-epoch data also show consistently an extended tail of weaker emission, of which G2 appears to be the head. Several possibilities have been proposed for the origin of the cloud: Gillessen et al. (2012) showed that a simple test particle model matches the observations well. Burkert et al. (2012) and Schartmann et al. (2012) used hydrodynamic simulations of pressure confined gas clouds with originally spherical or shell-like geometries to match the data. The idea of a gas shell was also put forward by Meyer & Meyer-Hofmeister (2012), who propose that a recent nova outburst could have ejected a ring-like shell of gas. Also a stellar source might hide inside G2: It might be a young planetary nebula (Gillessen et al. 2012), an evaporating protoplanetary disk (Murray-Clay & Loeb 2012), or it might even be the result of a collision between a giant and a stellar-mass black hole in the central cusp (Miralda-Escude 2012). Morris, Meyer & Ghez (2012) briefly review these various ideas. Here, we report the results of our 2012 observations, updating the orbital parameters and presenting new, deep spectroscopy.

## 2. Observations & data reduction

On March 14, 2012 we obtained an additional epoch of L-band imaging using the adaptive optics imager NACO on the VLT (Lenzen et al. 1998, Rousset et al. 1998). From the 50 frames with an integration time of 30 seconds each, 30 passed the quality cut. The frames underwent the standard procedure for data reduction. The astrometry was derived in our usual scheme: From the co-add we determined a point spread function using the starfinder code (Diolaiti et al. 2000), and the image then was Lucy-deconvolved (Lucy 1974). G2 is clearly visible (figure 1), although it is confused with some other source, which might be the dust feature reported by Clenet et al. (2005). Hence, we assigned the position a larger error of 5 mas per axis (for the other years we estimated an error of 2 mas). The astrometric frame we defined very locally by means of the surrounding S-stars, for which we took the positions from Gillessen et al. (2009).

From the end of March to mid July 2012 we obtained deep integral field spectroscopy with SINFONI on the VLT (Eisenhauer et al. 2003, Bonnet et al. 2004, table 1). We used the combined H+K grating (R=1500), and an image scale of 12.5 mas/pix. A total of 810 minutes on-source integration was achieved, of which 670 minutes pass our quality cut, demanding that a star in the field of view has a FWHM size of less than 7 pixels. We applied the standard data reduction for SINFONI, omitting however the division by a dedicated telluric standard star. Instead, we used the spectrum of the B2 dwarf S2 from the data themselves for that, minimizing the systematics related to airmass and atmospheric changes. We removed hydrogen and helium absorption lines in the spectrum of S2 by local Gaussian fits. After spatial registration and co-addition of the individual cubes, G2 stands out clearly as a highly redshifted component in Br-γ (2.166 $\mu$m), He I (2.058 $\mu$m) and Pa-α (1.875 $\mu$m). The latter line is located between the H- and K-bands and therefore harder to detect from ground-based data despite the fact that the line intrinsically is brighter than the other two K-band lines. G2 is unconfused in the emission lines, even at its location in the central, most crowded 200 mas. We created three combinations: The March epoch, the July epoch and a combination of all 2012 data. In the combined cubes, S2 has a FWHM of 5.7 pixels. From the two shorter cubes we determined the radial velocities and the positions in March and July. The combined cube we used to extract a position-velocity-diagram. To that end, we projected the updated, best-fitting orbit into the cube and extracted the spectrum along this virtual, curved slit (figure 2). We then rebinned the position-velocity diagrams to the same velocity scale around the three lines such that we were able to co-add them. The same technique of curved slits and co-addition of the lines we also applied to the data sets from 2011, 2008 and 2004 (except that Pa-α was blended too much with atmospheric lines for the 2004 data). We calibrated the line fluxes with the continuum between 2.17 $\mu$m and 2.18 $\mu$m of S2, for which we assumed $m_K$ = 14 (Gillessen et al. 2009) and an extinction of $A_K$ = 2.42 (Fritz et al. 2011).

## 3. Results

*3.1 Update of the orbit: increased eccentricity*
We re-fitted the orbit including the 2012 astrometry and radial velocities (figure 3), using the same gravitational potential as before from Gillessen et al. (2009). Table 2 compares the orbital elements from Gillessen et al. (2012) and the new fit. The new data confirm the conclusion of a highly eccentric orbit coplanar with the young stellar disk (Paumard et al. 2006, Lu et al. 2009, Bartko et al. 2009). The orientation of the orbit has not changed significantly. The addition of the 2012 data yields an eccentricity of 0.966, even higher than estimated from the previous data only. As a result the epoch of pericenter is two months later than found in Gillessen et al. (2012), around September 10, 2013. G2 will plunge deeper into the potential, with a nominal pericenter distance of 2200 Schwarzschild radii $R_S$ only. The previous best fit yielded 3100 $R_S$ (figure 4). This is bound to increase the importance of hydrodynamic effects during the pericenter passage, since the density profile of the hot gas around Sgr A* is expected to increase roughly like $r^{-1}$ (Xu et al. 2006). The higher eccentricity and the other changes are mainly due to the high radial velocities we have measured: The part of the radial velocity curve currently sampled around the maximum value is particularly sensitive to the eccentricity (figure 5).

*3.2 Position-velocity diagrams: a tail along the orbit*
In figure 6 we show the position-velocity diagrams for 2004, 2008, 2011, and 2012. The diagrams show impressively how G2 has accelerated while moving closer to Sgr A*. Also, the growing velocity gradient is apparent. Figure 7 is a three-color composite of the diagrams from 2008, 2011 and 2012. In 2012, the line width has increased to 600 km/s FWHM. This large increase in width compared to 2008 (230 km/s) and 2011 (370 km/s) is intrinsic and is not caused by the co-addition of spectra taken at different times in spring 2012. The FWHM is at most 20 km/s smaller in the July combination than in the full 2012 data cube, consistent with the measured (and expected) change in $v_{LSR}$ of +100km/s between March and July 2012.

The 2012 data is the deepest of the four epochs available. The tail of emission already seen in Gillessen et al. 2012 is now detected at high significance. We can trace the tail to the Southeast out to the point where it gets confused with the minispiral emission (Paumard, Maillard & Morris 2004), at the edge of the field of view. A faint bridge connects head and tail in the position-velocity-diagram. Interestingly, the combination of updating the orbit and using a virtual curved slit (instead of a simple longslit) has put the tail and bridge right onto the orbit. Another noteworthy feature is that there is a precursor - some weak emission from gas in front of G2 at even higher velocities - up to more than 2500 km/s.

*3.3 Flux estimates: G2 did not change flux*
While the line width of the head has increased, the surface brightness has decreased in a way that the total line flux remained constant within the uncertainties during all four years at $(2.0 \pm 0.2) \times 10^{-3}$ $L_\odot$. The 2012 data is of good enough quality to also estimate the flux across the full structure. For the flux estimation, we separate tail, bridge, head and precursor in velocity space. We obtain for the tail ranging from +242 to 696 km/s a total flux of $(2.1 \pm 0.3) \times 10^{-3}$ $L_\odot$, comparable to the flux in the head, but with larger systematic uncertainty, because the measurement might be affected by confusion with the ambient Br–γ gas. Also, the tail is spatially extended much more than the head with a length of at least 400 mas, and the tail appears to be clumpy (figure 8). The bridge (796 – 1488 km/s) has a luminosity of $(0.4 \pm 0.1) \times 10^{-3}$ $L_\odot$, similar to the precursor (2249 – 2803 km/s) for which we get $(0.3 \pm 0.1) \times 10^{-3}$ $L_\odot$.

With our new data, we can also estimate the mass in the tail, following the same calculation as done in Gillessen et al. (2012) for the head. The scaling of the mass of a gas cloud in case B recombination theory is $M \sim f_V^{1/2} V^{1/2} L^{1/2}$, where $f_V$ is the volume filling factor, V the emitting volume and L the luminosity. Inspecting figure 8, it seems justified to assume that a beam-smeared filament of 100mas length can approximate the tail's volume, which is thus roughly four times the volume of the head. The clumpy appearance means that $f_V < 1$. Since the luminosities of head and tail are equal, an estimated value of $f_V = 0.25$ makes the masses of head and tail comparable, while $f_V = 1$ would mean that the tail has twice the mass of the head.

**4. Discussion**

*4.1 Large eccentricity*
Perhaps the most puzzling finding is the high value for the eccentricity. A value close to 1 means that the orbit is almost fully radial - which is an unlikely configuration in phase space. For a relaxed distribution of orbits, the eccentricity distribution is $n(e) \sim e$ (Schödel et al. 2003). In such a system the probability to find an orbit with $e \geq 0.966$ is 6.7%. Hence, one can ask, what led to this high value. One possibility is that the cloud initially had a much less eccentric orbit, and that interaction with the hot ambient gas has erased the tangential component of the initial velocity. The gravitational force of the black hole will then tend to dominate over time the direction of the motion. In this picture, the large value of the eccentricity thus is a sign of the interaction of G2 with the ambient gas.

Another possibility is that G2 was created with almost no tangential velocity. The wind speeds of massive stars are comparable to the orbital velocities in the stellar disk, such that the material released from the backside of a massive star might happen to have low angular momentum. Also, the collision of two (or more) stellar winds might remove angular momentum from the gas. Furthermore, if G2 has formed as a result of a cooling instability in the accretion flow (section 4.2), a large eccentricity would be expected.

# New observations of the gas cloud G2 in the Galactic Center

More speculative is the idea that G2 originates from a radial outflow from the central black hole, naturally having low angular momentum. Such outflows or jets can occur in different accretion models (Falcke & Markoff 2000, Hawley & Balbus 2002). Interestingly, in the Southeast direction (from which G2 approaches Sgr A*), there is a 2.6-arcsecond long dust feature, visible in L-band images of the Galactic Center (figure 9). It is stationary in our data between 2002 and 2012, and approximately along the same direction is the outer orbit of G2. Beyond the morphology we have no information whether the dust feature is related to G2 or not, nor whether it might originate from Sgr A*. Its direction is also different from the proposed jet direction by Muzic et al. (2010) and Yusef-Zadeh et al. (2012). In this picture it is not clear why the orbit of G2 is coplanar with the stellar disk.

*4.2 Purely Keplerian motion*

The gas with velocities larger than 2500 km/s in the 2012 data can be explained by tidal interaction only. In figure 10 we show a test particle simulation, using the same cloud parameters as in Gillessen et al. 2012 (a Gaussian shape of the cloud in space with a FWHM of 25 mas, and a Gaussian velocity distribution with a FWHM of 120 km/s), but updating the orbit to the new estimate. One can see that by 2012 a fraction of the particles has reached velocities of $\approx$ 2500 km/s, very similar to what is seen in the position-velocity-diagram in figure 6. These particles are the ones, which happen to have the smallest pericenter distances. They occur roughly to equals parts due to the spatial and the velocity width of the simulated cloud. Using widths five times smaller does not yield the high-velocity gas anymore. The simulation also predicts that the first parts of the head have already reached pericenter, but the bulk will occur in 2013. The width of the pericenter epoch distribution is significant: It has a Gaussian FWHM of 1.3 years

The panels in the right column of figure 6 show the residuals of the position-velocity-diagrams after subtracting the best fitting orbit. The tail and the bridge appear to match the orbit fairly well. The simulations by Schartmann et al. (2012) on the other hand showed that hydrodynamic effects would lead to gas at velocities slower than the original orbit, i.e. one would expect negative velocities in the residuals. We conclude that the features visible in the position-velocity-diagrams are still consistent with purely Keplerian motion, and that no hydrodynamic effects are observed yet. Future simulations will show, how that finding starts to constrain the density profile of the accretion flow or the volume filling factor $f_V$ of the cloud.

The value for the eccentricity given by Gillessen et al. (2012) formally deviates by $3\sigma$ from the updated value, or by slightly less than $2\sigma$ after rescaling the errors such that the reduced $\chi^2$ of the fits is 1. Despite the low significance, one might speculate whether there is a reason for the increase in eccentricity. It might be explained in physical terms, if G2 would have lost angular momentum during the last year, and thus the change would point towards interaction with the hot ambient gas. However, we are not confident with such a claim, because our astrometric data points are few, and they are probably suspect to unrecognized confusion. Addition of new data can thus change the fit results more than captured in the formal fit errors.

*4.3 On the nature of G2*

The evolution of the Br-$\gamma$ luminosity is interesting in the light of the nature of G2. For the model of Murray-Clay & Loeb (2012), the amount of gas lifted from the protoplanetary disk will increase as G2 approaches the massive black hole, and the Br-$\gamma$ luminosity should increase. From their model the authors estimate an increase by almost a factor of two between mid 2011 and mid 2012. The fact that the measured Br-$\gamma$ luminosity has not changed from 2004 to 2012 argues against the evaporating protoplanetary disk. For a homogeneous gas cloud, the situation is more complicated. The luminosity is proportional to the electron density squared, and hence the tidal effects might affect the luminosity. There are two competing effects: Along the orbit the cloud gets stretched, and in the direction perpendicular it gets compressed. In addition, the surrounding gas might contribute to that compression. Also, if the cloud consists of clumps confined by an external force, the luminosity is less affected. The density inside the clumps would not change much, while the clumps get tidally sheared apart.

# New observations of the gas cloud G2 in the Galactic Center

The evolution of the size, line width and velocity gradient of G2 between 2004 and 2012 are well matched by an empirical model in which G2 formed as a compact spherical cloud relatively recently (between 1995 and 2000, Gillessen et al. 2012, Burkert et al. 2012). This means that the cloud would have formed well inside its apocenter distance, possibly due to wind-wind collisions (Cuadra et al. 2006). Such an origin seems somewhat unlikely, especially since it is then unclear why the angular momentum vector of G2 should be parallel to that of the clockwise disks of massive stars at radii larger than 1". Other models allowing an earlier formation, preferably at a radius close to the apocenter distance, would a priori seem to be more likely.

Burkert et al. 2012 noticed that a gas cloud starting in pressure equilibrium at apocenter of the given orbit should have been stretched into an almost linear, spaghetti-like feature long before it had a chance to get near Sgr A*. This does not match the head of G2 at first glance, which is much more compact. However, this stretching might be a natural explanation for the tail that (with the updated orbit) perfectly follows the trajectory. Head, bridge, and tail of G2 are a spatially extended, almost linear structure. The head would then just be an overdensity at the leading part of the whole gaseous structure, perhaps recently created by a local interaction in order to allow for its observed compactness. Also the shell models introduced by Schartmann et al. (2012) and Meyer & Meyer-Hofmeister (2012) remain an interesting option, in particular since we estimated that the mass in the tail is comparable to the mass in the head of G2, as long as it is possible to form locally compact structures during the inward orbital motion. Yet, in the shell model presented by Schartmann et al. (2012), the tail does not follow the same orbit as the head, unlike what the new observations show.

A scenario that can explain (i) the large eccentricity, (ii) the ballistic motion without evidence for interaction with ambient gas even only one year before pericenter passage, and (iii) the non-negligible internal velocity dispersion needed for the test particle model is the following: G2 formed by a cooling instability from the ambient gas. An overdense region of hot gas when going through a cooling instability would break up into little droplets of cold gas with a velocity dispersion equal to the initial sound speed of the region. Noticeably, the internal velocity dispersion of $\approx 100$ km/s matches in this sense the speed of sound for gas with temperatures of order a few million Kelvin, as present in the accretion flow. In between those droplets there is still diffuse hot gas. So the droplets might be pressure confined by hot inter-droplet gas. As a result of buoyancy, this system of droplets will begin to fall inwards towards the SMBH on a radial orbit. If the mass and radius of the droplets are small, they will not be tidally torn apart. In addition, the system, will follow a ballistic trajectory, like the cluster of test particles in the simulation.

Eckart et al. (2012) argue that G2 might rather be a dusty stellar object, for example a narrow-line Wolf-Rayet star. By the faintness of G2 this already seemed unlikely given the Gillessen et al. (2012) data, but the 2012 velocity width of the Br-γ line of 600 km/s in the new SINFONI data rules out that explanation now. Furthermore, Eckart et al. (2012) present detections of G2 in K- and H-band. A marginal detection in K-band is consistent with the spectral energy distribution presented by Gillessen et al. (2012), according to which G2 should have $m_K \approx 18$ (with at least half a magnitude of uncertainty). The H-band detection on the other hand would require a stellar source in G2. We note, however, that the H-band positions reported by Eckart et al. (2012) are coincident within the uncertainties with the $m_H \approx 19.5$ star S63 (Gillessen et al. 2009, figure 11) that is moving in a direction similar to the one of G2. The H-band flux thus cannot be ascribed to G2 alone, if this identification is justified at all. The H-K colors derived by Eckart et al. (2012) also match a stellar source.

*4.4 Future observations*
The shift of the epoch of the pericenter to a later time might look less favorable for ground-based near-infrared observations, since the Galactic Center is observable with the VLT from late February to early October each year only. However, it is not clear that the infrared wavebands are the ones reacting most quickly to the increased mass inflow. The material still has to spiral down to few $R_S$, a process which probably is dominated by the viscous time scale of $\approx 100$ years (Mościbrodzka et al. 2012). Hence, even if only a fraction of the time scale is needed, the flux increase still might be many months after the nominal pericenter date. Furthermore, the intrinsic spread of the pericenter epochs will smear out the event over more than a year. More direct chances of observing the impact have X-ray and radio observatories (Gillessen et al. 2012, Burkert et al. 2012, Narayan, Özel & Sironi 2012), both of which have longer observability windows compared to the near-infrared. Figure 12 shows the evolution of the X-ray luminosity as predicted from the shock model and RIAF model used in Gillessen et al. (2012), but for the updated orbit. According to this model, in spring 2013 the X-ray luminosity of the shock will start to be brighter than the quiescent luminosity of Sgr A* (Baganoff et al. 2003). During pericenter passage, the flux increase would reach a factor 30. Due to the small spatial scales, for X-ray observatories Sgr A* and G2 are confused.

## 5. Summary

We have presented new data for the gas cloud G2 currently falling towards Sgr A*. Our main conclusions are:

- We confirm that G2 moves on a highly elliptical orbit in the plane of the clockwise stellar disk.
- Our new observations confirm the prediction of accelerated tidal disruption of a diffuse gas cloud inferred from data between 2004 and 2011; the speed and degree of tidal disruption is larger than predicted from these earlier data, however, because the eccentricity of the orbit of G2 is larger (e = 0.966).
- The predicted epoch of pericenter passage has shifted to early September 2013, and the nominal pericenter distance has decreased to 2200 $R_S$ only.
- G2 consists of a comparably compact head (with a velocity shear of 600 km/s by now) and a more widespread tail of at least 400 mas length.
- The observed dynamics is fully consistent with purely Keplerian motion, i.e. we don't detect any hydrodynamic effects.
- The near unity eccentricity of the orbit is highly puzzling, because in a random distribution of orbits the value of 0.966 would only have a probability of 6.7%, inconsistent with Occam's razor.
- The flux of G2 has not changed between 2004 and 2012, making an evaporating protoplanetary disk scenario unlikely.

Overall, a simple ballistic gas cloud, which is being tidally sheared, is an excellent description of the data available. The origin of such a cloud remains puzzling. The case for an exciting and scientifically fruitful impact of G2 onto Sgr A* has been strengthened.

# New observations of the gas cloud G2 in the Galactic Center

# New observations of the gas cloud G2 in the Galactic Center

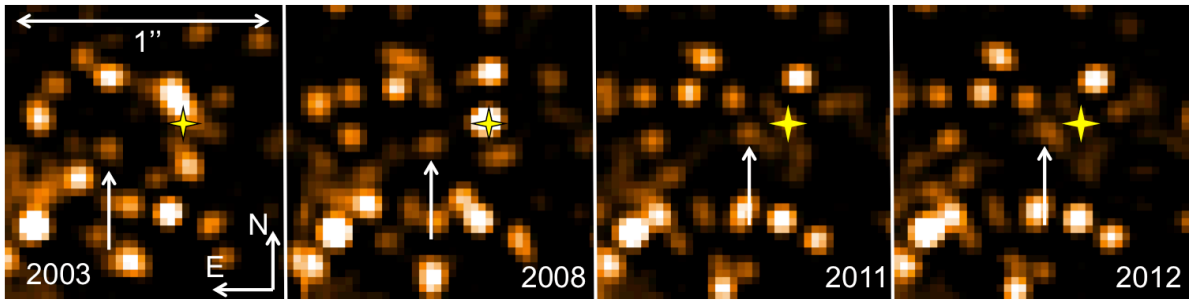

Figure 1: Sequence of L'-band images of the central arcsecond. In each frame the cross marks the position of Sgr A*, which was actually flaring during the 2008 exposure. The arrow marks G2, the inward motion of which is obvious.

Table 1: Summary of the SINFONI data collected in 2012 on G2.

| Date | Minutes on-source | Minutes after quality cut | Data used in |
|---|---|---|---|
| March 18 | 40 | 40 | March & 2012 combination |
| May 5 | 40 | 40 | 2012 combination |
| May 20 | 40 | 40 | 2012 combination |
| June 25 | 30 | 30 | July & 2012 combination |
| June 29 | 210 | 180 | July & 2012 combination |
| June 30 | 80 | 50 | July & 2012 combination |
| July 6 | 170 | 90 | July & 2012 combination |
| July 7 | 200 | 200 | July & 2012 combination |
| **Total** | **810** | **670** | |

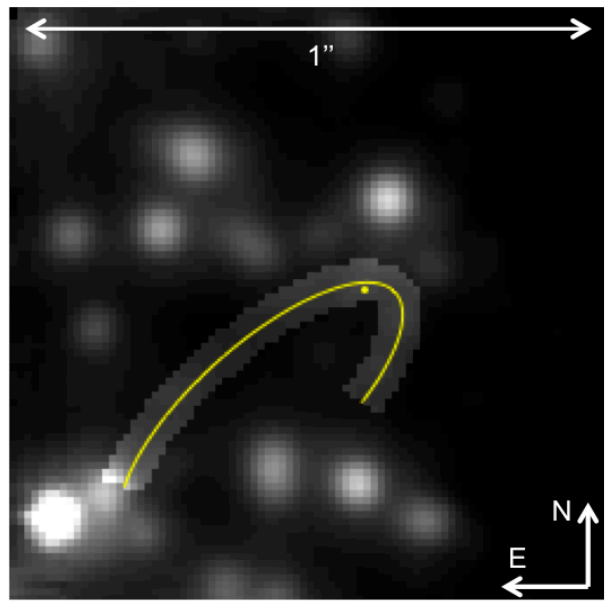

Figure 2: The projected best-fitting orbit of G2 overlaid on an image of the S-star cluster from integrating the SINFONI 2012 data around Br-γ. Sgr A* is marked by the dot. The curved slit used for extracting the position-velocity-diagram is indicated by the gray shaded area.

# New observations of the gas cloud G2 in the Galactic Center

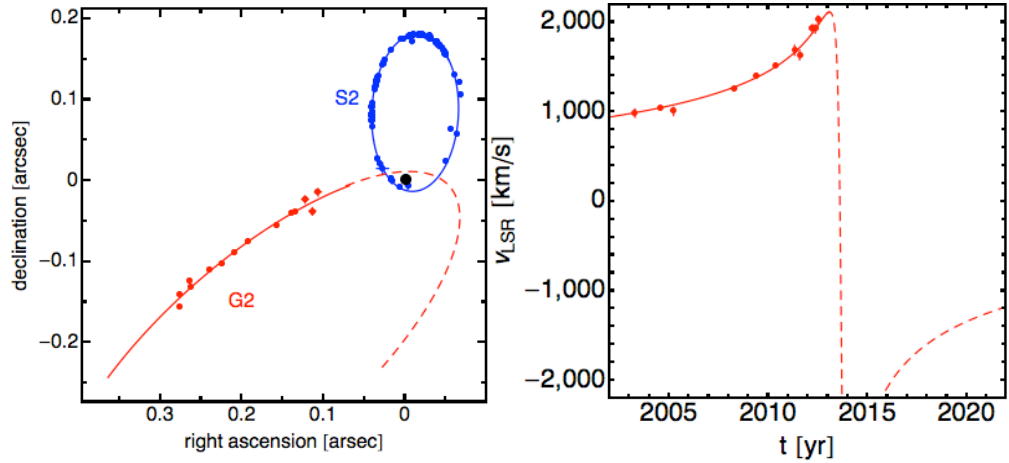

Figure 3: Orbital data and best fitting orbit for G2 (in red). Left: The astrometric data (points) and the best-fitting orbit. The 2012 data (at α = 0.1" / δ = 0") suffers from higher positional noise than the previous years. Right: Radial velocity (LSR corrected) of G2. Since 2011, the radial velocity has increased by 400 km/s. The error bars are plotted, but barely larger than the plot symbol size.

Table 2: Orbital parameters for the orbit of G2.

|  | Gillessen et al. 2012 | Updated fit |
|---|---|---|
| semi major axis (mas) | 521 ± 28 | 666 ± 39 |
| eccentricity | 0.9384 ± 0.0066 | 0.9664 ± 0.0026 |
| inclination [°] | 106.55 ± 0.88 | 109.48 ± 0.81 |
| position angle of ascending node [°] | 101.5 ± 1.1 | 95.8 ± 1.1 |
| longitude of periastron [°] | 109.59 ± 0.78 | 108.50 ± 0.74 |
| epoch of periastron [yr] | 2013.51 ± 0.04 | 2013.69 ± 0.04 |
| orbital period [yr] | 137 ± 11 | 198 ± 18 |

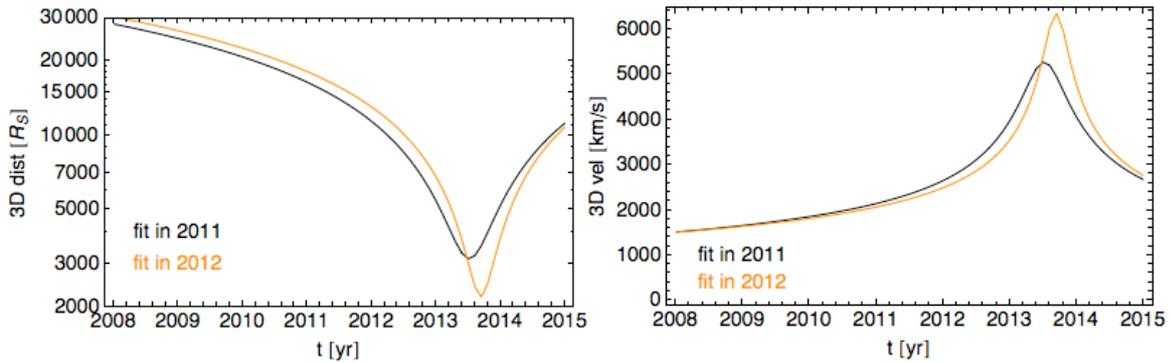

Figure 4: Comparison of the pericenter passages for the orbit given in Gillessen et al. 2009 (black) and for the updated orbit (orange). Left: 3D distance in Schwarzschild radii; right: 3D velocity in km/s.

# New observations of the gas cloud G2 in the Galactic Center

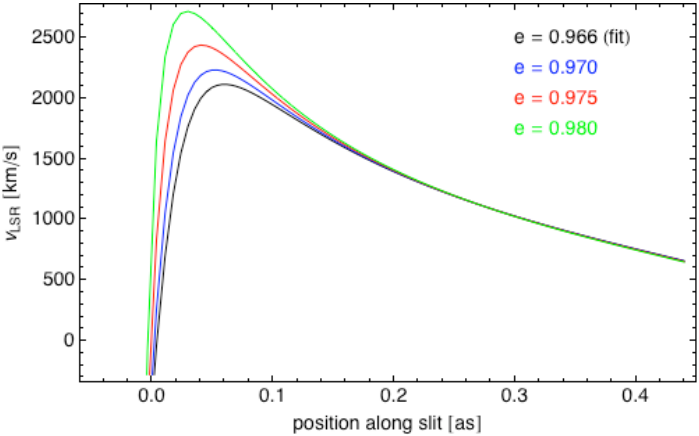

Figure 5: Position-velocity-diagram of different orbits. The black curve is for the best-fitting orbit; the others are obtained by changing the eccentricity and keeping all other parameters fix. The maximum radial velocity is very sensitive to the eccentricity.

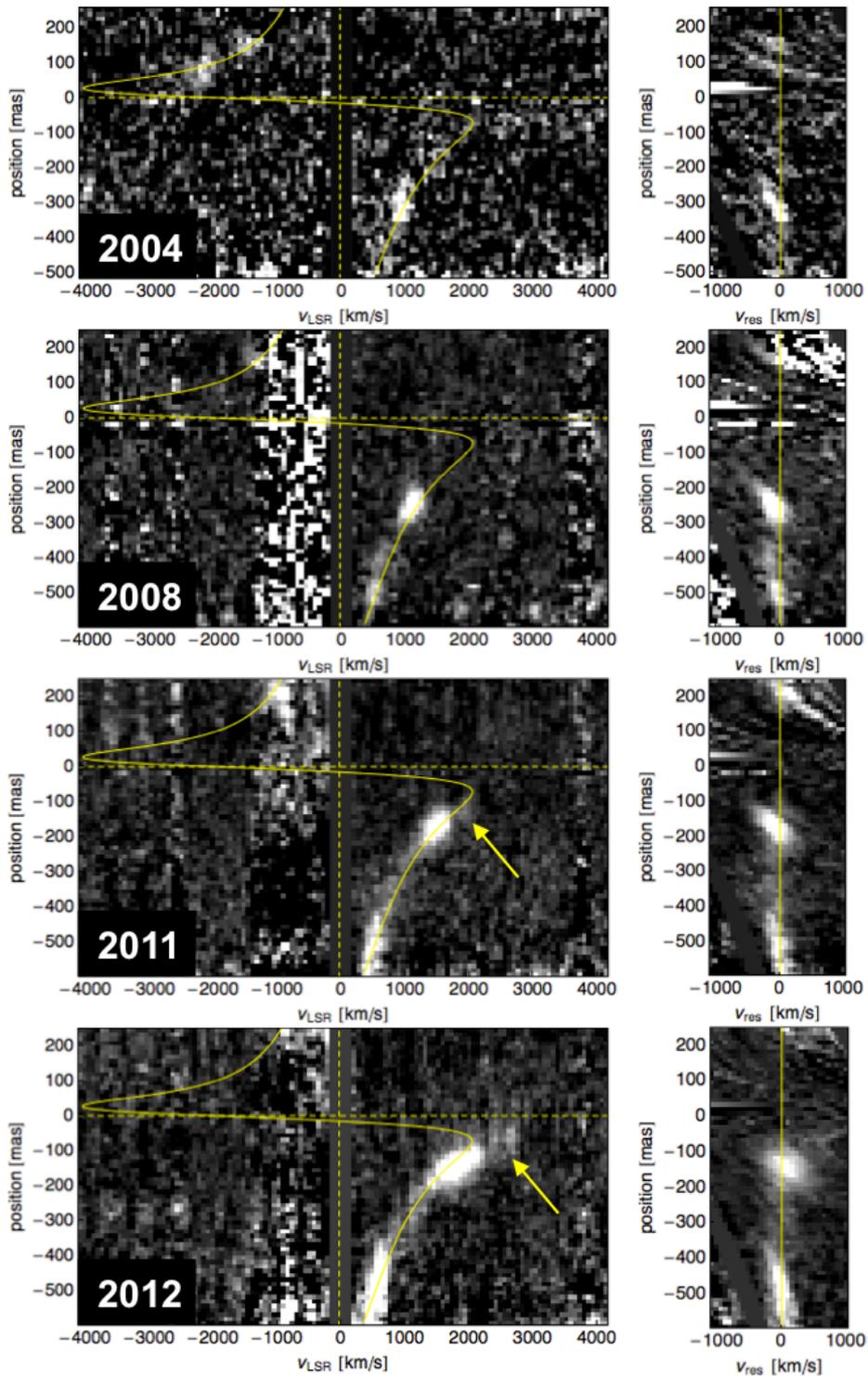

Figure 6: Position-velocity diagrams of G2, extracted along the virtual curved slit as shown in figure 2 and explained in the text. Left column: The solid, yellow line denotes the best-fitting orbit. The dashed lines mark 0 velocity and pericenter passage. The yellow arrows mark the precursing gas. Right column: Residuals after subtracting the best fitting orbit.

New observations of the gas cloud G2 in the Galactic Center

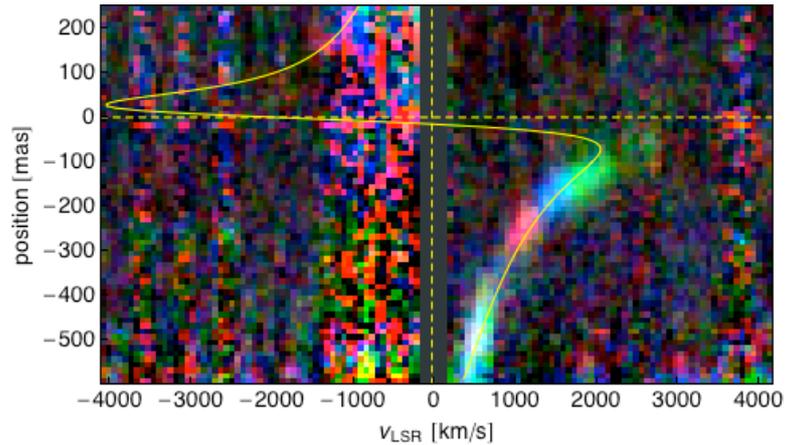

Figure 7: Three-color overlay of the position-velocity diagrams for 2008, 2011 and 2012 as shown in figure 6.

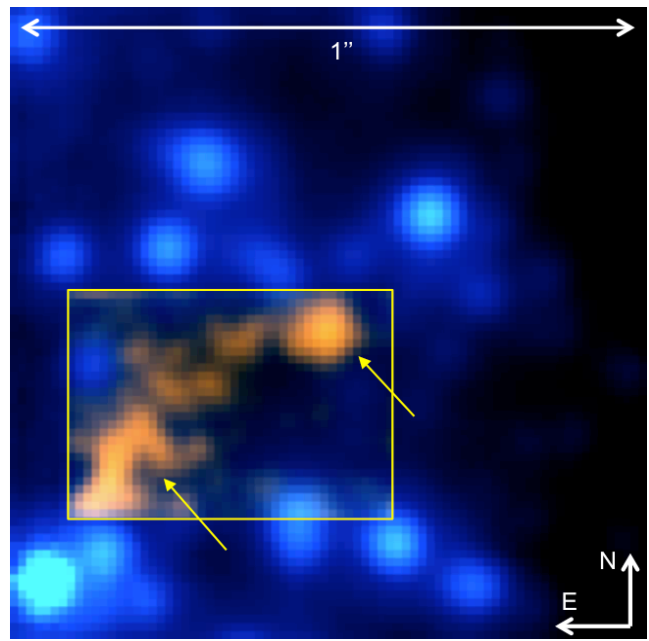

Figure 8: Br-γ map of G2. The image was obtained from the 2012 SINFONI data set by extracting 25 channel maps from 2.1695 $\mu$m to 2.1815 $\mu$m in the yellow rectangle, each with a spectral smoothing of 5 pixels and subtracting the spectrally neighboring channels. In each channel map then a σ-clipping at σ = 1.5 was applied. Finally, for each pixel the brightest value of the 25 channel maps was selected (since the gas at different positions is at different velocities), and the image was smoothed with a Gaussian beam with FWHM=3 pix. The head and the tail are marked with arrows.

New observations of the gas cloud G2 in the Galactic Center

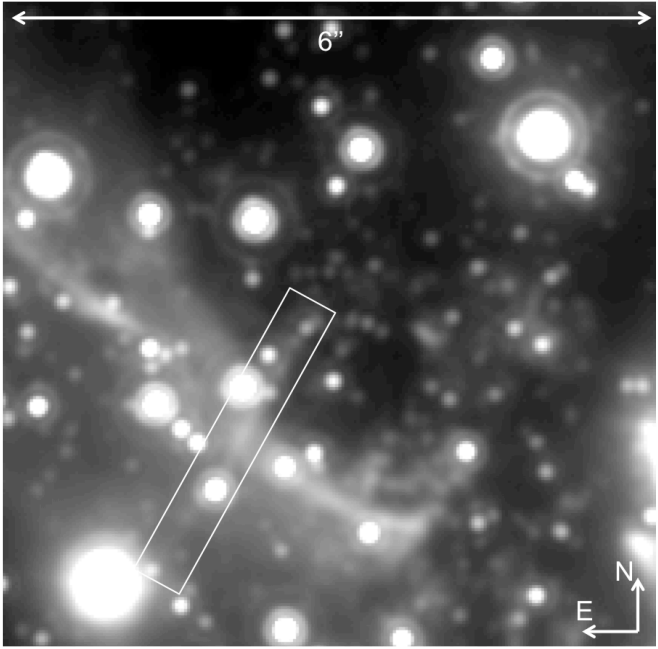

Figure 9: L'-band (3.8 $\mu$m) image of the Galactic Center obtained with NACO at the VLT on July 6, 2011. Note the 2.6"-long dust feature to the Southeast, roughly along the direction from which G2 approached Sgr A*. The brighter gas streamer from the Northeast towards the Southwest is the so-called mini-spiral.

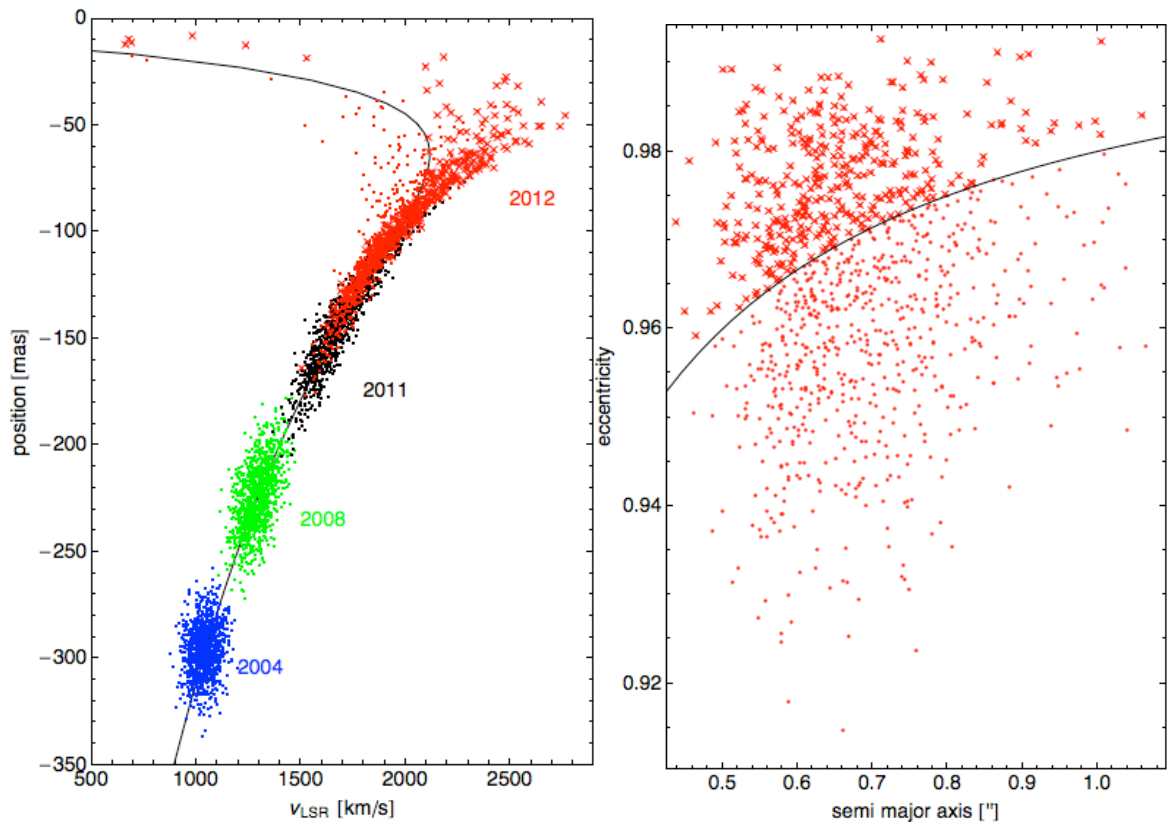

Figure 10: Test particle simulation of the gas cloud. Left: During the four epochs, G2 develops a strong velocity gradient, and in 2012 a tail to velocities as high as 2500 km/s can be seen. Right: Distribution of semi-major axis and eccentricity for the test particles. The black line corresponds to a pericenter distance of 20 mas. In both panels the crosses are for those particles that have a pericenter distance less than 20 mas. The high velocity tail in the left panel is made of these particles.

New observations of the gas cloud G2 in the Galactic Center

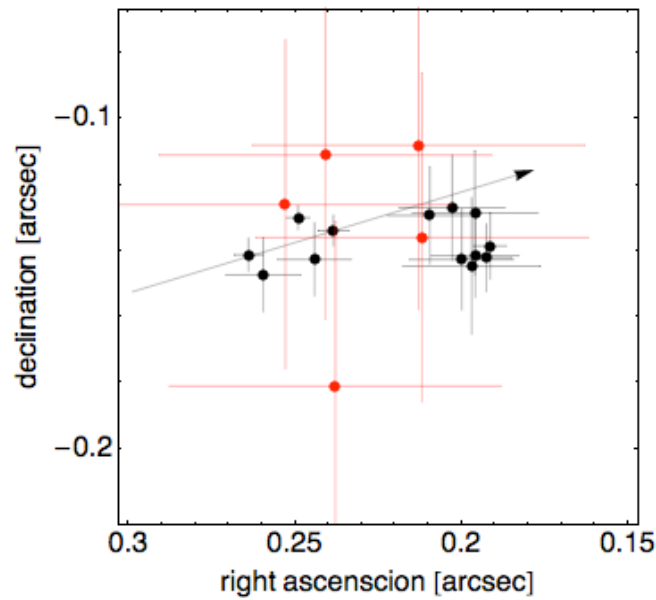

Figure 11: The red data are the five H-band positions of G2 reported by Eckart et al. (2012), assuming a positional error of 50 mas. The black data are H-band positions for the star S63 (Gillessen et al. 2009) between 2004 and 2009. The errors are scaled such that the combined K-band and H-band data set (going up to 2012) yields a reduced $\chi^2$ of 1 when assuming a parabolic law of motion. The black line represents that fit. S63 is moving from the Southeast to the Northwest, like G2.

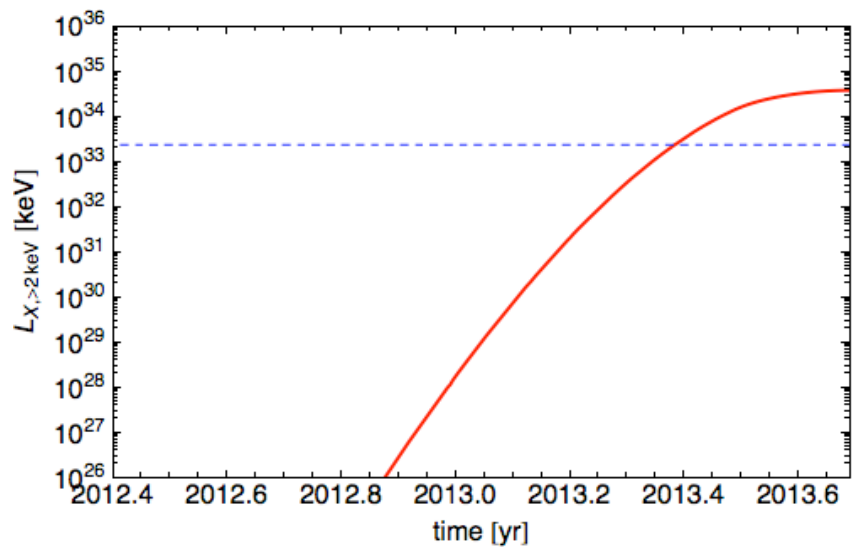

Figure 12: Predicted evolution of the X-ray luminosity above 2 keV for the shock model used in Gillessen et al. 2012 and the updated orbit.